\documentclass[review]{elsarticle}

\usepackage{hyperref}
\journal{Journal of \LaTeX\ Templates}

\usepackage{pst-all}
\usepackage{amssymb}
\usepackage{amsthm,mathptmx}
\usepackage{amssymb}

\newtheorem{Proof}{Proof}

\newtheorem{Definition}{Definition}
\newcounter{example}%[section]
\newenvironment{example}[1][]{\refstepcounter{example}\par\medskip
   \noindent \textbf{Example~\theexample. #1} \rmfamily}{\medskip}
\usepackage{graphicx}
\usepackage{float} 
\usepackage{appendix}
\usepackage{multirow,array}
\usepackage{array,multirow,makecell}
\setcellgapes{1pt}
\makegapedcells
\newcolumntype{R}[1]{>{\raggedleft\arraybackslash }b{#1}}
\newcolumntype{L}[1]{>{\raggedright\arraybackslash }b{#1}}
\newcolumntype{C}[1]{>{\centering\arraybackslash }b{#1}}
\usepackage{booktabs}
\usepackage{caption, booktabs, tabularx, makecell}

\usepackage{algorithm}
\usepackage{algorithmic}
\makeatother
\restylefloat{algorithm}
\usepackage{array}
\usepackage{pdflscape}
\begin{document}

\begin{frontmatter}

\title{Obsolete  Personal Information Update System for the Prevention of Falls among Elderly Patients}


\author[add1,add4]{Salma Chaieb\corref{mycorrespondingauthor}}
\cortext[mycorrespondingauthor]{Corresponding author}
\ead{salma.chaieb2@yahoo.com}

\author[add2,add4]{Brahim Hnich}
\ead{hnich.brahim@gmail.com}

\author[add3,add4]{Ali Ben Mrad}
\ead{benmradali2@gmail.com}

\address[add1]{University of Sousse, ISITCom, 4011, Sousse, Tunisia;}
\address[add2]{University of Monastir, FSM, 5000, Monastir, Tunisia;}
\address[add3]{University of Sfax, ISAAS, 1013, Sfax, Tunisia;}
\address[add4]{University of Sfax, CES Lab, 3038, Sfax, Tunisia;}

\begin{abstract}
Falls are a common problem affecting the older adults and a major public health issue. Centers for Disease Control and Prevention, and World Health Organization report that one in three adults over the age of 65 and half of the adults over 80 fall each year. 
In recent years, an ever-increasing range of applications have been developed to help deliver more effective falls prevention interventions. All these applications rely on a huge  elderly personal database collected from hospitals, mutual health, and other organizations in caring for elderly. 
The information describing an elderly is continually evolving and may become obsolete at a given moment and contradict what we already know on the same person. So, it needs to be continuously checked and updated in order to restore the database consistency and then provide better service.
This paper provides an outline of an Obsolete personal Information Update System (OIUS) designed in the context of the elderly-fall prevention project. Our OIUS aims to control and update in real-time the information acquired about each older adult, provide on-demand consistent information and supply tailored interventions to caregivers and fall-risk patients. The approach outlined for this purpose is based on a polynomial-time algorithm build on top of a causal Bayesian network representing the elderly data. The result is given as a recommendation tree with some accuracy level.
We conduct a thorough empirical study for such a model on an elderly personal information base. Experiments confirm the viability and effectiveness of our OIUS.
\end{abstract}


\end{frontmatter}

%\linenumbers

\section{Introduction}

The sharp increase in the life expectancy of the world’s population results in a large number of older adults. This progressive aging has enormous social and economic consequences, which will prove to be crucial in the coming decades. Falls are the main factor contributing to this rise. They are more common in older adults around the world and may have several painful consequences leading in worst cases to death. A large part of the population is at risk. According to the Centers for Disease Control and Prevention, around $30\%$ of people over $65$ have at least one serious fall with adverse effects per year. Compounding this is population aging. Indeed, at 80 years old, over half of seniors fall annually.
Falls can be the leading cause of admission to hospital for older adults. As the incidence of hospital admissions increases exponentially with age, the social and health costs associated with falls are expected to increase. Falls, with or without injury, can also affect the quality of life, social relationships, function, and mental health of the elderly. A growing number of older adults fear to fall and, as a result, limit their activities and social engagements. These impacts also lead to enormous economic losses.
Statistics have shown that these fall-related costs could be significantly reduced by using preventive training strategies. Thus, the development of automatic monitoring and fall-prevention systems that call for help from caregivers such as lifestyle guides, clinical-community partnerships, etc. becomes a serious necessity.
In this respect, several fall prevention strategies and tools are being developed and tested involving treating physicians, other health care team members, older adults, and some of their family members \cite{varshney:mobile,chan:effectiveness}.

To adequately meet the needs, all these systems require monitoring elderly personal information base collected from hospitals, mutual health, and other associations and organizations in caring for the elderly.
Information is usually provided by widely distributed sources and is often uncertain and unreliable. Furthermore, it is likely to be changed as time goes by and may become, at some time, obsolete and contradict other information. So, it needs to be consistently updated upon the acquisition of new observations about older adults, in order to maintain and retrieve consistent real-time information on-demand.
However, the permanent collection of personal information and the constant monitoring of the quality of the information held is a costly and time-consuming process. Therefore, in this paper, we opt for an intelligent and autonomous agent to meet this need.

An approach has been proposed in \cite{cbhd:data} to maintain database consistency by detecting obsolete information that contradict a newly  acquired certain event. The approach in \cite{cbhd:data} is based on the assumption of the existence of a causal Bayesian network (BN) that encodes the relationships among the features in the database. Obsolete information that contradict a newly acquired certain event is then detected, in \cite{cbhd:data}, using a polynomial-time algorithm exploiting the causal BN. Such obsolete information is then presented, with a certain confidence, in the form of the so-called AND-OR tree to describe the possible logical ways the existing information can be in contradiction with the new event.

In this paper, building on the work in \cite{cbhd:data},  we propose a decision support system to manage information  obsolescence of a database of older adults gathered during the elderly appointments with their attending physician in both the University Hospital Center of Lille and Valenciennes (France). Such a decision support system will help practitioners by providing two types of recommendations: (1) detect potential obsolete information about the elderly as new events are recorded and recommend most probable substitute values; and (2)  predict most probable values for the elderly when such information is missing from the database. 
To this end, our contributions in the paper can be summarized as follows:

\begin{itemize}

\item We design a causal BN modeling the general knowledge about elderly adults through a laborious process that
ends by adopting the BN by the domain experts;

\item We design a recommendation system and show the usefulness of this system in detecting obsolete information as newly certain events arrive; and
    
\item We validate the quality of our recommendations by conducting intensive simulations using historical medical information of older adults.
\end{itemize}

The rest of the paper is structured as follows: In section \ref{section2}, we introduce the overall system architecture. In Section \ref{section3}, we describe the construction details of the BN for preventing falls in older adults. In section \ref{section4}, we give the overall obsolete information update system architecture.
In Section \ref{section5}, we present our empirical results. Finally, in Section \ref{section6}, we draw some final conclusions and point out directions for future work.

\section{Overall System Architecture}
\label{section2}

In this section, we first present the general context and motivate the work. Then, we outline the overall architecture of the proposed decision-support system.

\subsection{Context and motivation}
%\label{section2}

Technological advances are placing increasing importance on elderly monitoring and have pushed the frontier of healthcare into the home settings \cite{liu:mobile,kandakoglu:decision, nasir:decision}.
The prevention of falls requires a pedagogical and educational approach in which the attending physicians are privileged actors. In the context of the shortage of physicians and work overload, physicians reported working on average between 40 and 60 hours per week, according to the 2014 Work/Life Profiles of Today’s Physician\footnote{https://www.ama-assn.org/practice-management/physician-health/how-many-hours-are-average-physician-workweek}. A fall prevention support system for physicians can really help improve this educational work. To be relevant, such a system should require very little time from the physician during a consultation.
It must avoid the long and tedious collection of information necessary to assess the risk of an elderly person falling and provide appropriate recommendations.

The general context of the presented work is the Elderly-Fall Prevention project which is part of the ELSAT2020\footnote{http://www.elsat2020.org/en} project.
It also relates to the "sustainable mobility and handicap" axis within the Polytechnic University of Hauts-de-France and is in collaboration with the university hospitals of Valenciennes and Lille.

Throughout this paper, the personal database attributes are denoted with uppercase letters such as $X_1$, $X_2$, $X_3$. The domain of an attribute $X_i$ is denoted with $\mathcal{D}(X_i)$. Specific values, also called observations, taken by those attributes are denoted with lowercase letters $x_{1,1}$, $x_{2,1}$, $x_{3,2}$ with $x_{i,j} \in \mathcal{D}(X_i)$, $j \in \{1, ..., \#\mathcal{D}(X_i)\}$. 
Let $o_{new}$ denote the newly observed value of a variable $O_{new}$ such that $(O_{new}, o_{new})$ was not in $\textbf{OBS}$ at the previous iteration. Let $\textbf{OBS'}=\textbf{OBS} \setminus (O_{new}, o_{new})$.

The objective, in this paper, is to design an intelligent and autonomous system to control and update in real-time the information $\textbf{OBS}$ acquired about each older adult. 
For each older adult and before updating the database row reserved for him with the newly acquired observation, we identify which, among his existing observations contained in $\textbf{OBS'}$, are now obsolete and contradict the new observation.
Proper recommendations are then provided to the users (e.g., doctors) on how to update the database row reserved for the concerned person. 

The motivation is to have an information base that includes as much information as possible on the population in question, which is consistent and continually updated.
Having such an information base can contribute to the improvement of fall prevention. It is, in fact, an innovative and effective way to help clinicians, nursing staff, mentors, elder's relatives, follow the elderly by providing them with consistent information about the target person.
This can help detect the early warning signs of elderly falls and protect them from falls.
It also allows them to share consistent information, provide proactive support and collaborate in fall prevention.

\subsection{Overall Architecture}

\begin{figure}[h]
  %\centering
  \hspace{-1cm}
  \includegraphics[scale=0.7]{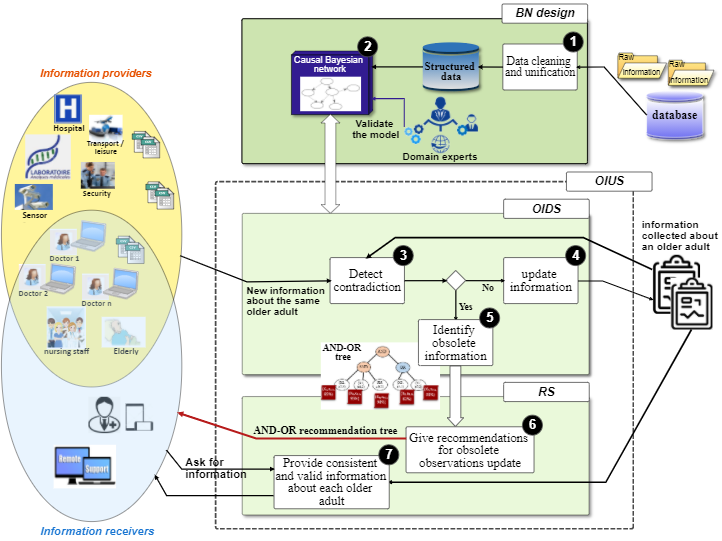}
  \caption{Obsolete Information Update System.}
  \label{fig2:system}
\end{figure}

Fig. \ref{fig2:system} shows the overall architecture of our system. The first module, \textit{BN design}, is responsible for building the causal BN representing our knowledge about the elderly. The result obtained is a BN validated by some domain experts. 
Once the BN is created, we use it to monitor the behavior of the concerned elderly people in real-time in order to detect and prevent the risk of falling. 

Fig. \ref{fig2:system} also shows the overall architecture of our Obsolete Information Update System (\textit{OIUS}) responsible for continuously tracking and monitoring the information about a specific subject and recommending the possible obsolete observation to be updated. Our OIUS is composed of two sub-modules; the Obsolete Information Detection System (\textit{OIDS}) and the Recommender System (\textit{RS}). The Obsolete Information Detection System (OIDS), which uses the BN to approximately detect contradictions between observations and identify possible obsolete ones. In case of contradiction, the result obtained from the OIDS is an AND-OR tree, which is used by the Recommender System (RS) to identify the most likely observations to be updated. The RS identifies the minimum, but sufficient, number of questions to be asked of users in order to replace the obsolete observations and expand the database.

Let us consider the following scenario example shown in Fig. \ref{fig1:exp} illustrated by using some part of the causal BN:

%figure parts of the BN with var and obs that illustrate the example. Try to find example relates to the fall prevention
\begin{figure}[h]
  \centering
  \includegraphics[scale=0.8]{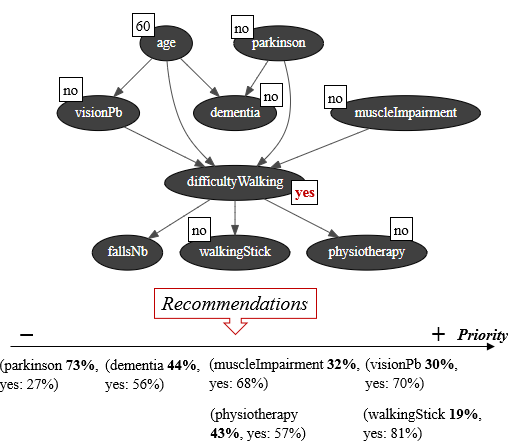}
  \caption{A scenario example on some part of the causal BN.}
  \label{fig1:exp}
\end{figure}

\begin{example}
\label{example1}
At time $t_1$, the set of observations \textbf{OBS} includes seven observations related to an older adult: Mr. Wilson is \textit{60 years old}, he has \textit{good eyesight}, he has \textit{no muscle impairment}, he has \textit{no Parkinson disease}, he does not suffer from \textit{dementia}, and thus he \textit{doesn't use a walking stick} and \textit{is not receiving physiotherapy}.

At time $t_2$, new information, that is supposed to be certain, indicates that: Ms. Wilson \textit{has a serious walking problem}.
\end{example}

For our elder, the arrival of the new observation raises a contradiction since it seems that there is no reason to have walking disorders. Our system tries then to identify among the  observations of \textbf{OBS'} those that may be obsolete.
First, it detects that Mr. Wilson may have visual disturbances, may be suffering from muscle impairment, dementia, or possibly Parkinson's disease that may cause difficulty walking.
In order to assist decision-makers, we try to prioritize the possible obsolete observations. So, we attribute to each observation a coefficient based on some probability measures and taking into account the acquisition time of each observation. 
%We assume that the observations with the lowest coefficients are likely to be updated. In the case where two observations have equal coefficients, the older one will be processed first. 
We suggest also the most likely values that can replace the obsolete ones with some prediction accuracy. As Mr. Wilson hasn't checked his vision for a long time, we try to ask him about his sight: \textit{"Mr. Wilson, have you noticed the deterioration of your eyesight?"}. He confirmed his good eyesight. At this time, and since dementia usually develops after the age of $70$ and about $10$ to $15$ years after diagnosis of Parkinson's disease (general knowledge), we suggest to his general practitioner, after processing the new information, that he may have a muscle impairment.
%and we recommend that he refer him to a neurologist. Our system can also provide the specialist doctor with additional information about the older adult, helping to better understand Mr. Wilson's state of health. 
Second, our system detects that the non-use of a cane or physical therapy can exacerbate the problem. So, 
as shown in Fig. \ref{fig1:exp} 
it can help the doctor by providing recommendations on the possible use of a walking stick \textit{(walkingStick, yes, 81\%)} or the need for physiotherapy \textit{(physiotherapy, yes, 57\%)} to prevent falls.
As shown in Fig. \ref{fig1:exp}, 
The choice of observations to be updated follows a certain priority. We will show later how to choose which obsolete observations to update, how to assign priorities to obsolete observations, and how to calculate predictive values that can replace obsolete ones to help users with decision making.

%%%%% when users choose from the list of obsolete observations the ones to be deleted and when he gives the new value, we start a new cycle

\section{Causal Bayesian network design for preventing falls in older adults}
\label{section3}

In this section, we describe the process of building our causal BN model for preventing falls in older adults. 

\subsection{BN Background}

Probabilistic graphical models, such as Bayesian networks \cite{pearl:bayesian, jensen:intro, darwiche:modeling}, offer a coherent representation of domain knowledge under uncertainty. They are based on the solid foundations of probability theory and they readily combine available statistics with expert judgment. BNs are successfully applied to a variety of problems, including machine diagnosis, natural language interpretation, vision, data mining, and they have also been proven effective in medicine. Examples include kidney transplant survival prediction \cite{topuz:predicting}, multiple-disease risk assessment \cite{wang:directed}, healthcare access management by predicting patient no-show behaviour \cite{ferro:improving}, etc.

%In this work, we use capital letters such as $X_1$ , $X_2$ , $X_3$ to denote variable names and lowercase letters such as $x_{1,1}$, $x_{2,1}$, $x_{3,2}$ for specific values, also called observations, taken by those variables. We denote by $N$ the number of BN variables.

A BN is a couple $(G, \Theta)$ that consists of a qualitative part, encoding causal relationships among a domain's variables $\textbf{X} = \{X_1,...,X_N\}$ in a directed graph $G$, and a quantitative part, encoding the joint probability distribution $\Theta$, given by the chain rule (\ref{eq1}), over these variables. It is defined by the product of the local probability distributions on each variable $X_i$, $P(X_i \mid Pa(X_i))$, where $Pa(X_i)$ denotes the set of the parents of the node $X_i$ in $G$.
\begin{equation}
    P (\textbf{X}) = \prod_{i=1}^{n} P (X_i \mid \emph{Pa}(X_i))
    \label{eq1}
\end{equation}

Each node $X_i$ of the graph represents a random variable and each arc represents a direct dependence between two variables. Indirect influences are represented by paths through the network.
Formally, the structure of the directed graph $G$ is a representation of a factorization of the joint probability distribution. As many factorizations are possible, there are many graphs capable of encoding the same joint probability distribution. Of these, those that reflect the causal structure of the domain are especially convenient and preferred. They are best used for analyzing causal problems as they reflect the expert's understanding of the domain, improve interaction with a human expert at the stage of model building, and are easily scalable with new information.
Every independence between two variables leads to omitting an arc from the graph. This leads therefore to significantly reduces the number of arcs needed to fully quantify the domain.
The precise relationship between probabilistic independence and connectedness of nodes is defined in terms of a property called d-separation \cite{gvp:d-separation}.
The quantification of a BN is to assign prior probability distributions to variables that have no predecessors in the network and conditional probability distributions to variables that have predecessors. Thus, a Conditional Probabilities Table (CPT) is associated with each variable. These probabilities can easily incorporate available statistics and, where no data are available, expert judgment.
The most important reasoning type in BNs is known as belief updating or inference and consists of calculating the probability distribution over variables of interest given other observed variables, i.e. the probability distribution over the model variables is adjusted for a particular case, in which given values are assigned to some model variables. Efficient inference procedures exist for performing computations over the
network \cite{lauritzen:local}.

\subsection{Model Structure}
As part of the Elderly-Fall Prevention project, we have access to a real-life database $\emph{Elderly-Data}$ that contains information on the elderly. It is collected during the elderly appointments with their attending physician in both the University Hospital Center of Lille and Valenciennes over a 9-year period (2005-2014). The $\emph{Elderly-Data}$ includes about $1174$ patient records, each of these records was described by $207$ patient-history features (binary, denoting presence or absence of a feature or continuous, expressing the value of a feature).
The construction of the Bayesian structure is conducted using the following steps;

Step 1: by reviewing literature published in the field of elderly falls and by analyzing the database $\emph{Elderly-Data}$, we identified $41$ features out of $207$ that represent the most significant direct or indirect influences of falls among older adults. In \cite{sharif:falls}, authors have reported several risk factors that may increase elderly falls and that include history of falling, environmental hazards such as poor lightening, and various health conditions including muscle weakness, visual and hearing disorders, balance impairments, cognitive and sensory impairments, diabetes mellitus, orthostatic hypotension, and osteoporosis. Several studies have also associated some drugs with an increased risk of falls among older adults. The most common ones are different types of psychotropic drugs and  cardiovascular drugs \cite{early:joint}.
The number of drugs taken by the older adult is also a predictor for falls as the medications interactions can cause side effects and increase the risk of falls in older adults \cite{montero:polypharmacy}.

In addition to the bibliographic research, we analyzed the attributes of the database and applied some statistical measures in order to have an overview of the most frequent factors of falls. Table \ref{tab1} summarizes the frequencies of the significant features. 
%More than half $(60.3\%)$ of participants who have fallen at least once are in the age group of 73-81 years.
The analysis shows that people with falls tended to have sensory deficits ($61\%$) and motor disorders ($69\%$).
The rate of falls increased with age so that for those 64-72 years $36.5\%$ had experienced falls and for those 73-81 years $60.3\%$ had. 
Analysis in table \ref{tab1} also shows that the health status of older adults and the diseases from which they suffer greatly influence the risk of falls ($533$, $58.4\%$).
The order of health consequences of the falls was fracture ($533$, $56.4\%$) and fear falling ($521$, $55.2\%$).
Once the variables were selected, we asked the experts to validate them. They checked if other factors, considered as direct influencers, can be added or if certain variables need to be removed. This step ended with the validation of the $41$ selected variables.

\begin{table}[!h]
\begin{tabular}{|p{2.5cm}|r|c|c|c|c|}
\hline
\multirow{2}{=}{\thead{Items}}                & \multicolumn{1}{c|}{\multirow{2}{*}{\textbf{Values}}} & \multicolumn{2}{c|}{\textbf{Frequency (\%)}} & \multirow{2}{*}{\textbf{Total}} & \multirow{2}{*}{\textbf{\begin{tabular}[c]{@{}c@{}}\%\\ n = 1174\end{tabular}}} \\ \cline{3-4}
                                                  & \multicolumn{1}{c|}{}                                 & Fallers              & Non-fallers           &                                 &                                                                                 \\ \hline
\multirow{2}{=}{Gender}                           & female                                                & 481 (56.4)           & 371 (43.6)            & 852                             & 72.5                                                                            \\
                                                  & male                                                  & 117 (36.4)           & 205 (63.6)            & 322                             & 27.5                                                                            \\ \hline
\multirow{4}{*}{Age}                              & 55 - 63                                               & 4 (21.1)             & 15 (78.9)             & 19                              & 1.6                                                                             \\
                                                  & 64 - 72                                               & 43 (36.5)            & 75 (63.5)             & 118                             & 10.1                                                                            \\
                                                  & 73 - 81                                               & 280 (60.3)           & 184 (39.7)            & 464                             & 39.5                                                                            \\
                                                  & $\geq 82$                                          & 350 (61.1)           & 223 (38.9)            & 573                             & 48.8                                                                            \\ \hline
\multirow{2}{*}{Living situation}                 & living alone                                          & 366 (59)             & 256 (41)              & 622                             & 53                                                                              \\
                                                  & living with family                                    & 317 (57.4)           & 235 (42.6)            & 552                             & 47                                                                              \\ \hline
\multirow{6}{*}{Causes of falls}                  & chronic pathology                                     & 139 (39)             & 217 (61)              & 356                             & 30.3                                                                            \\
                                                  & sensory deficits                                      & 235 (61)             & 151 (39)              & 386                             & 64.3                                                                            \\
                                                  & motor disorders                                       & 605 (69)             & 271 (31)              & 876                             & 51.5                                                                            \\
                                                  & diseases                                              & 683 (58.4)           & 485 (41.5)            & 1168                            & 58.2                                                                            \\
                                                  & drugs interference                                    & 569 (48.8)           & 599 (51.2)            & 1168                            & 48.4                                                                            \\
                                                  & environment                                           & 399 (58.5)           & 282 (41.5)            & 681                             & 34                                                                              \\ \hline
\multirow{4}{*}{Falls number}                     & 0                                                     & \multicolumn{2}{l|}{}                        & 230                             & 19.6                                                                            \\
                                                  & 1 - 2                                                  & \multicolumn{2}{l|}{}                        & 457                             & 39                                                                              \\
                                                  & 3 - 4                                                 & \multicolumn{2}{l|}{}                        & 236                             & 20.1                                                                            \\
                                                  & $\geq 5$                      & \multicolumn{2}{l|}{}                        & 251                             & 21.3                                                                            \\ \hline
\multicolumn{1}{|l|}{}                            & \multicolumn{1}{l|}{}                                 & \multicolumn{3}{c|}{\textbf{Total}}                                            & \textbf{\begin{tabular}[c]{@{}c@{}}\%\\ n = 944\end{tabular}}                   \\ \hline
\multirow{2}{=}{Consequences of falls} & fracture                                              & \multicolumn{3}{c|}{533}                                                       & 56.4                                                                            \\ \cline{2-6} 
                                                  & fear falling                                          & \multicolumn{3}{c|}{521}                                                       & 55.2                                                                            \\ \hline
\end{tabular}
\caption{The number of falls in the elderly and their causes and health consequences.}
\label{tab1}
\end{table}

\begin{figure}[h]
  \centering
  \includegraphics[scale=0.7]{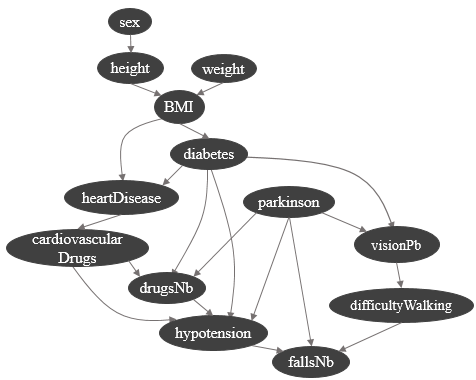}
  \caption{A first BN with 13 variables.}
  \label{fig:RB13}
\end{figure}

Step 2: at this stage, we have used the selected variables to build the BN structure.
From bibliographic searches in international databases, peer-reviewed medical literature, and health statistic reports we have built up a documentary collection of $200$ references relating to the selected features and which justify relationships between them. We have used these references and the experts' opinions to elicit the BN structure. We started by building a first model with $13$ variables as shown in Fig. \ref{fig:RB13}. Then, we have gradually extended the first model by adding variables one at a time.
In \cite{sharif:falls}, authors state that cardiovascular drugs such as diuretics and beta-blockers may cause or worsen orthostatic hypotension and falls. It has also been stressed by the same authors that poly-pharmacy and the use of psychotropic drugs, especially when combined with cardiovascular medications increase the risk of falls in the elderly. Fall risk is proved to be closely related to severe cognitive impairment in elderly individuals who have Parkinson \cite{chomiak:differentiating}. This justifies the following directed edges, in Fig. \ref{fig:RB13}: \emph{cardiovascularDrugs} $\rightarrow$ \emph{hypotension}, \emph{drugsNb} $\rightarrow$ \emph{hypotension}, \emph{hypotension} $\rightarrow$ \emph{fallsNb}, and \emph{parkinson} $\rightarrow$ \emph{fallsNb}.
Several other studies have also associated psychotropic drugs with an increased risk of falls among elderly as it can cause impaired balance and coordination \cite{bareis:association}. This finding allows us to extend the BN structure by adding the two variables \emph{psychotropicDrugs} and \emph{difficultyBalance} connected by an edge directed from the first to the second variable, and an edge directed from \emph{psychotropicDrugs} to \emph{drugsNb}.
Deterioration in biological systems, such as declines in sensory system capacity and motor functions causes delays in the stabilization of control systems, which could contribute to postural instability and falls \cite{jia:prevalence}. Recent epidemiological findings of associations between hearing capacity and motor functions have proven the strong association between hearing loss, poor balance, and falls \cite{carpenter:effects}.
Balance and gait ability is one of the primary factors for falls \cite{thomas:physical}. Gait disturbance in non-Parkinsonian elderly correlates is very closely correlated with the future emergence of Parkinson's disease \cite{galazky:deep}. Elderly with type 2 diabetes have significantly worse balance control and less efficient gait control possibly leading to higher fall risk \cite{deshpande:sensory}. Based on these findings, we extend the BN structure by adding the following directed edges as shown in Fig. \ref{BN}: 
%as shown in Fig. \ref{fig:ext1}:
\emph{hearingPb} $\rightarrow$ \emph{difficultyBalance},
\emph{difficultyBalance} $\rightarrow$ \emph{fallsNb},
%\emph{difficultyWalking} $\rightarrow$ \emph{fallsNb}.
\emph{Parkinson} $\rightarrow$ \emph{difficultyWalking},
\emph{Diabetes} $\rightarrow$ \emph{difficultyBalance}.
%\begin{figure}[h]
%  \centering
% \includegraphics[scale=0.7]{extension1.PNG}
%  \caption{A part of the BN structure.}
%  \label{fig:ext1}
%\end{figure}
%In \cite{ }people who fall experience fear of falling, which leads to further limiting of activity
The prevalence of falls is higher among people with stroke than those without. 
%Falls are a common consequence after stroke and may lead to fear of falling and reduced activity \cite{denissen:interventions}.
Among victims with stroke, elderly have a higher mortality rate and increased risk of disability. Many patients experience difficulty with movement, including balance issues and gait disturbances, after stroke \cite{choi:immediate}. Stroke may also result in decreased truck muscle strength and limited trunk coordination, frequently determining loss of autonomy due to the trunk impairment \cite{sorrentino:clinical}.
Factors associated with increased stroke risk in multivariate analysis included age, diabetes and cardiovascular disease \cite{sharma:stroke}.
%According to World Health Organization, forty percent of traumatic injuries-related hospitalizations are due to falls. The most common fall-related consequences are pain and fracture including upper extremity and hip fractures.
%In \cite{chiodini:falls}, authors indicated that there is high quality evidence that vitamin D plus calcium reduces the risk of falls and any type of fracture in particular osteoporotic fractures.
This finding allows us to add the variables \emph{age}, \emph{muscleImpairment}, \emph{autonomyLoss} and \emph{strokeTIA}, and the following directed edges:
\emph{strokeTIA} $\rightarrow$ \emph{fallsNb},
\emph{strokeTIA} $\rightarrow$ \emph{difficultyBalance},
\emph{strokeTIA} $\rightarrow$ \emph{difficultyWalking},
\emph{strokeTIA} $\rightarrow$ \emph{muscleImpairment},
\emph{strokeTIA} $\rightarrow$ \emph{autonomyLoss},
\emph{muscleImpairment} $\rightarrow$ \emph{autonomyLoss},
\emph{age} $\rightarrow$ \emph{strokeTIA},
\emph{diabetes} $\rightarrow$ \emph{strokeTIA},
\emph{heartDisease} $\rightarrow$ \emph{strokeTIA},
\emph{fallsNb} $\rightarrow$ \emph{fracture},
as shown in Fig \ref{BN}.
%Fig. \ref{fig:ext2} shows the extension part of the BN structure.
%\begin{figure}[h]
%  \centering
%  \includegraphics[scale=0.67]{extension2.PNG}
%  \caption{A part of the BN structure.}
%  \label{fig:ext2}
%\end{figure}
Whenever a directed edge is added between two variables, we ask for experts assistance. 
A total of 4 experts were interviewed to validate the structure of the BN and evaluate the relationships between the nodes.
We estimate that elicitation of the structure took about $36$ hours with the experts. This includes model refinement sessions, where the previously obtained structure was reassessed in a group setting. The structure of our current model is shown in Fig. \ref{BN}. We believe that it models reasonably causal interactions among the selected variables.
\begin{landscape}
\pagestyle{empty}
\begin{figure}[h]
\vspace{-5.5cm}
\hspace{-6cm}
 \includegraphics[scale=2.5]{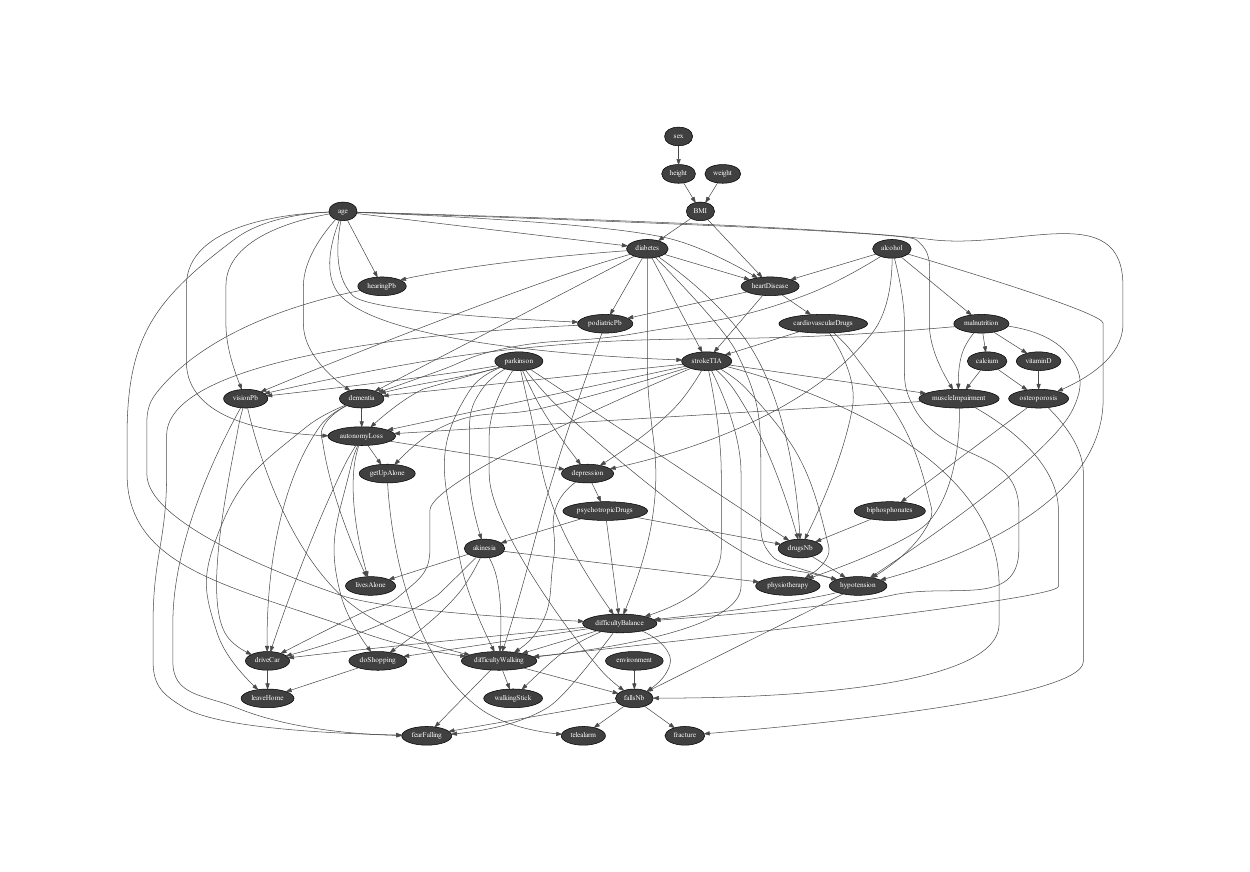}
 \vspace{-3cm}
 \caption{Causal BN for preventing falls among older adults.}
 \label{BN}
\end{figure}

\end{landscape}

\subsection{Model Parameters}
In order to draw useful information from the resulting BN, knowledge of the joint probability is required. 
The information in CPTs may be obtained from empirical data, direct measurements, related model outputs, or expert opinion under no other data available. 
In our work, and due to the lack of data, we essentially use experts' opinions to elicit the BN parameters, and we use data when it is possible.
Although the BN formalism allows both discrete and continuous variables, all exact general-purpose algorithms for BNs deal with models containing only discrete variables.
In order to take advantage of these algorithms, we first decided to discretize the continuous variables while accepting the minimum loss of variable information.
Our discretization concerns only $9$ variables and is based on expert opinion that variables such as \emph{driceCar}, \emph{doShopping}, or \emph{leaveHome} have essentially \emph{never}, \emph{less than once a week}, and \emph{at least once a week} values. Variable such as \emph{age}, \emph{drugsNumber}, and \emph{fallsNumber} are discretized with the method of \emph{Equal Frequency Discretization} \cite{tsai:optimal}.
It consists of transforming each continuous variable into a finite number of disjoint intervals with approximately the same distribution of data points, relates these intervals with meaningful labels, and re-encode, for all instances, each value on this attribute by associating it with its corresponding interval.
The numerical boundaries of these intervals are then checked and occasionally adjusted based on expert judgment.
Once the variables are discretized, we obtain a model with $41$ discrete variables; $32$ variables are binary, the two variables height and weight have respectively seven and ten states, and the other seven ones have between three and four states.

Given the structure of the model and the specification of the desired discretization, we used the database, available statistics, bibliographic searches, and solicit the experts to estimate the parameters of the model. Prior probability distributions are simply relative counts of various outcomes for each of the variables in question. However, although the prior probabilities can be learned reasonably accurately from the database of hundreds of records, the conditional probabilities present more of a challenge. In cases where several variables directly precede a variable in question, individual combinations of their values may be very unlikely to the point of being absent from the data file. In such cases, we assumed that the distribution is uniform and we solicited experts to validate it. Thus, for each of these variables, the simple average of distributions from each expert was used to produce the corresponding CPT for the final model.
Note that in all cases where the counts were zero and a naive interpretation would suggest zero probability, we have inserted a small probability reflecting that almost no probability in the uncertain world is exactly zero or equal to one.

In general, BN model validation is quite complicated due to the limited availability of validation data. In our study, the validation of the BN includes expert evaluation, given the unavailability of sufficient data. Fortunately, there is much anecdotal and some empirical evidence that imprecision in probabilities has only a small impact on the diagnostic accuracy of a system based on a BN \cite{phpdh:sensitivity}. Furthermore, one of the main advantages of a BN model is that it can be easily updated with new information, meaning it can be improved when new data or system knowledge become available.

\section{Obsolete Information Update System}
\label{section4}

In this section, we present the obsolete information detection system as well as our recommender system.

\subsection{Obsolete Information Detection System}

\begin{figure}[h]
  \centering
  \includegraphics[scale=0.62]{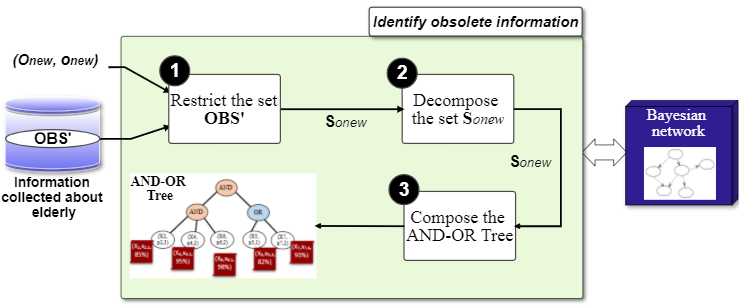}
  \caption{Obsolete information identification phase.}
  \label{fig4}
\end{figure}

Once the Bayesian model is created, we use it to monitor the behavior of the concerned elderly people in real-time in order to detect and prevent the risk of falling. An Obsolete Information Detection Algorithm (OIDA) was proposed by \cite{cbhd:data} to detect all possible contradictory observations, whenever new observations are acquired. 
In \cite{cbhd:data}, a contradiction between observations occurs when the conditional probability of the new observation given other observations is very close to 0. Since there is always a degree of uncertainty related to BN, the authors introduce a contradiction probability tolerance value $\epsilon$ to reflect those uncertainties related to the probabilistic dependencies among the variables in the BN. The approximate contradiction is thus defined as follows:

\begin{Definition}[\textbf{$\epsilon$-Contradiction \cite{cbhd:data}}]
\label{definition1}
Given a BN, a set of observed variables $\textbf{OBS'}$, a new observation $o_{new}$ on a variable $O_{new}$, and a real number $0 \leq \epsilon \leq 1$.

\hspace{-0.6cm} $\textbf{OBS'}$ is $\epsilon$-Contradictory   to $o_{new}$ when 
\[
P (O_{new} = o_{new}|\textbf{OBS'})
\leq \epsilon
\]
\end{Definition}
The obsolete information detection approach is based on this fundamental definition. Indeed, an $\epsilon$-Contradiction occurs when there is a subset $\textbf{S}_{o_{new}}$ of $\textbf{OBS'}$ of observations that have become obsolete and contradict $o_{new}$.
The process of identifying the set of obsolete observations $\textbf{S}_{o_{new}}$ takes place in 3 steps: (1) restrict the ${\epsilon}$-Contradictory set $\textbf{OBS'}$ into $\textbf{S}_{o_{new}}$; (2) decompose the ${\epsilon}$-Contradictory set $\textbf{S}_{o_{new}}$ looking for obsolete observations; and (3) compose the AND-OR tree from the set of obsolete observations $\textbf{S}_{o_{new}}$. Fig. \ref{fig4} summarizes these 3 steps.

\begin{figure}[h]
  \centering
  \includegraphics[scale=0.73]{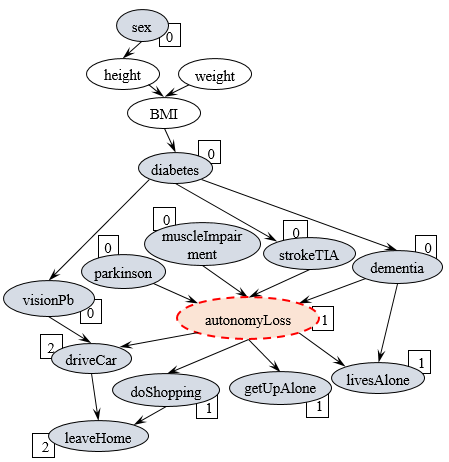}
  \caption{A scenario example on a part of the BN.}
  \label{fig5}
\end{figure}

Let us consider the example shown in Fig. \ref{fig5} on which we approach the three steps of the obsolete information identification phase.

\begin{example}
\label{example2}
At time $t_1$, the set of observations $\textbf{OBS}$ includes twelve observations related to an elderly person (Grey nodes in Fig. \ref{fig5}): she is a woman, she is not diabetic, she does not have dementia, does not have a stroke, does not have muscle impairment, and she does not suffer from Parkinson. Moreover, she has a good eyesight, drives her car regularly, does her shopping at least once a week, and leaves her home more than once a week. She can get up alone and she lives alone. 

At time $t_2$, new information (Dashed Node in Fig. \ref{fig5})  indicating that Ms. Wilson has lost her autonomy arrives.
\end{example}

\emph{\textbf{Step 1}}:
in \cite{cbhd:data}, authors show that the set $\textbf{OBS'}$ can be restricted to $O_{new}$'s dependent variables in $\textbf{S}_{o_{new}}$. 
For the example given in Fig. \ref{fig5}, the variables \emph{sex}, \emph{diabetes} and \emph{leaveHome} are conditionally independent of the variable \emph{autonomyLoss}, given the rest of the observed variables. So the set $\textbf{OBS'}$ is restricted in $\textbf{S}_{o_{new}}$ as follows:  $\textbf{S}_{o_{new}} = \{$ \emph{(dementia, 0), (strokeTIA, 0), (muscleImpairment, 0), (parkinson, 0), (visionPb, 0), (driveCar, 2), (doShopping, 1), (getUpAlone, 1), (livesAlone, 1)}$\}$.

\emph{\textbf{Step 2}}:
in \cite{cbhd:data}, authors decompose the set $\textbf{S}_{o_{new}}$ into subsets, $\textbf{S}_i$, by bringing the observations on dependent variables together. 
%Then, each subset $\textbf{S}_i$ is checked whether it is consistent with $o_{new}$ or not. If so, it will be ignored and will not be considered in the rest of the obsolete information identifying process. 
For the example given by Fig. \ref{fig5}, once \emph{autonomyLoss} is observed, its parents (\emph{dementia}, \emph{strokeTIA}, \emph{muscleImpairment}, \emph{parkinson}) are dependent,  and then they are grouped into a set $\textbf{S}_1$. We apply the same treatment on all the other variables to obtain four subsets of dependent variables as follows:
$\textbf{S}_1 = \{$\emph{(dementia,0), (strokeTIA,0), (muscleImpairment,0), (parkinson,0)}$\}$, $\textbf{S}_2 = \{$\emph{(dementia,0), (livesAlone,1)}$\}$, $\textbf{S}_3 = \{$\emph{(getUpAlone,1)}$\}$, and $\textbf{S}_4 = \{$\emph{(doShopping,1), (driveCar,2), (visionPb,0)}$\}$. 
%Then, in \cite{}, authors only keep the subsets that are contradictory to $o_{new}$. In this example, all the subsets are $\epsilon$-Contradictory to $o_{new}$, so, as a result of the first step of the decomposition phase we obtain the set $\textbf{S}_{o_{new}} = \{\textbf{S}_1, \textbf{S}_2, \textbf{S}_3, \textbf{S}_4 \}$. 
Then, following the approach proposed by \cite{cbhd:data}, each $\textbf{S}_i$ in $\textbf{S}_{o_{new}}$ is further divided into two disjoint ${\epsilon}$-Contradictory subsets: $\textbf{S}^{AND}_{i}$ and $\textbf{S}^{OR}_{i}$, such that: 
\begin{itemize}
  \begin{sloppypar}
    \item The \emph{AND-Set}, $\textbf{S}^{AND}_{i}$, contains each observation $x$ in $\textbf{S}_i$ that is individually ${\epsilon}$-Contradictory to $o_{new}$, given the BN, i.e., $P(O_{new} = o_{new} |x) \leq \epsilon$, and which is subject to update;
    \item The \emph{OR-Set}, $\textbf{S}^{OR}_{i}$, contains each observation that is not individually ${\epsilon}$-Contradictory to $o_{new}$, given the BN and which is likely to be involved in the contradiction.
    \end{sloppypar}
\end{itemize}

%All the observations in the \emph{AND-Set} are subject to update, which is not always the case for the observations in the set $\textbf{S}^{OR}_{i}$.

For the observations in the \emph{OR-Set}, authors in \cite{cbhd:data} state that with the available knowledge at their disposal, they are not able to accurately infer which one(s) should be updated. But they claim that each of these observations may be obsolete.
%The subset $\textbf{S}_1$ illustrate this delicate situation. It contains four observations: \emph{(dementia, 0)}, \emph{(strokeTIA, 0)}, \emph{(muscleImpairment, 0)} and \emph{(parkinson, 0)}. We suppose that the loss of autonomy is only caused by these four factors. $\textbf{S}_1$ is $\epsilon$-Contradictory to $(autonomyLoss, 1)$. However, if we take a closer look at each of these observations separately, we find that each of them is not individually $\epsilon$-Contradictory to $(autonomyLoss, 1)$. Indeed, since these observations are conditionally dependent,  $P(autonomyLoss=1 | dementia=0)>\epsilon$ comes from the fact that another factor among the three remaining led to the loss of autonomy. The same applies to the other observations. So, updating one of these four old observations is enough to remove the $\epsilon$-Contradiction. However, with the available knowledge at our disposal, we are not able to accurately infer which one(s) should be updated.
%\begin{sloppypar}
As a result of the decomposition phase, and for the given example, we obtain the set $\textbf{S}_{o_{new}} = \{\{\textbf{S}^{AND}_{1}, \textbf{S}^{OR}_{1}\}, ..., \{\textbf{S}^{AND}_{4}, \textbf{S}^{OR}_{4}\}\}$ of all possible obsolete observations, such that: 
%\end{sloppypar}

$\textbf{S}^{AND}_{1}= \{\emptyset\}$, $\textbf{S}^{OR}_{1} = \{$\emph{(dementia,0), (strokeTIA,0), (muscleImpairment,0), (parkinson,0)}$\}$

$\textbf{S}^{AND}_{2}= \{$\emph{(livesAlone,1)}$\}$, $\textbf{S}^{OR}_{2} = \{\emptyset\}$

$\textbf{S}^{AND}_{3}= \{$\emph{(getUpAlone,1)}$\}$, $\textbf{S}^{OR}_{3} = \{\emptyset\}$

$\textbf{S}^{AND}_{4}= \{$\emph{(doShopping,1), (driveCar,2)}$\}$,
$\textbf{S}^{OR}_{4} = \{\emptyset\}$

%\subsubsection{Compose the AND-OR tree from the set of obsolete observations}\label{compose}

\begin{figure}[H]
  \centering
  \includegraphics[scale=0.65]{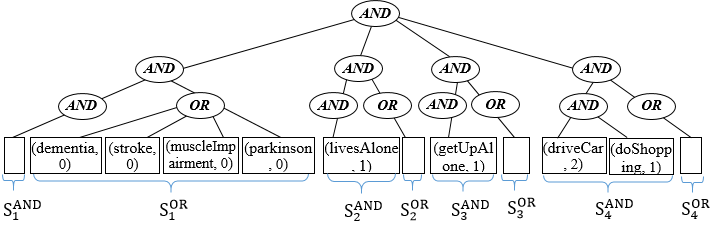}
  \caption{AND-OR tree associated with example \ref{example2}.}
  \label{fig6}
\end{figure}

\emph{\textbf{Step 3}}: 
the main aim of this step is to combine the results of the subsets of $\textbf{S}_{o_{new}}$ to create an AND-OR tree as follows: the root node is labeled \emph{AND}. Then, for each subset $\textbf{S}_i$, an \emph{AND} node whose parent is the root node is created. Next, for each \emph{AND-Set} (resp. \emph{OR-Set}) of $\textbf{S}_i$, an \emph{AND} (resp. \emph{OR}) node whose parent is the corresponding node of $\textbf{S}_i$, is added. Finally, a child leaf node for each observation in $\textbf{S}^{AND}_{i}$ (resp. $\textbf{S}^{OR}_{i}$) is created. Each leaf node is labeled with the obsolete observation.
The AND-OR tree represents precisely the set of obsolete observations $\textbf{S}_{o_{new}}$ and describes the logical relationships among its \emph{AND-Set}s and \emph{OR-Set}s. 
Fig. \ref{fig6} shows the AND-OR tree associated with the given example.
Indeed, given a new observation $o_{new}$ and an AND-OR tree $\mathcal{T}$ that represents the set of obsolete observations relative to $o_{new}$, the following three propositions, in \cite{cbhd:data},  are true:

\begin{itemize}
    \item For each observation $x \in \textbf{OBS'}$, $x \notin \mathcal{T}$ if and only if $x$ is not obsolete.
\item All observations of the \emph{AND-Set} are obsolete.
\item At least one observation of the \emph{OR-Set} is obsolete.
\end{itemize}

To better understand the full details of the obsolete information detection process, we refer the author to \cite{cbhd:data}.

\subsection{Recommender System}
The Recommender System (RS) aims to monitor and assess the risk of falls in older adults and to provide tailored recommendations and interventions to their caregivers.
The RS system offers two services. One takes place in case of contradictions and consists in recommending the most likely observations to be removed and giving most likely values that can replace them (step 6 in Fig. \ref{fig2:system}), while the other is consists in providing the caregivers with most likely information about  about their elderly patients in cases where such information is not available in the database (step 7 in Fig. \ref{fig2:system}).

\subsubsection{Removal and substitution recommendations}
First, in case of a contradiction, and based on the AND-OR tree provided by the OIDS, the RS tries to assess the set of all possible observations given by the resulting tree and select the most likely ones to be updated. For \emph{AND-Sets}, our RS recommends updating all the observations contained therein since they have certainly become obsolete. However, it gets a little more complicated when it comes to \emph{OR-Sets}. Indeed any observation in the \emph{OR-Set} may be obsolete but we are not able to exactly infer which one(s) should be updated. In this case, we propose a strategy based on studying the \emph{posterior} probability of the observations in the tree given $o_{new}$ to recommend which ones are the most likely ones to be removed.

Algorithm \ref{algo1} shows the main steps of the proposed method.
The inputs to the Obsolete Observations Recommendation Algorithm (OORA) are the AND-OR tree $\mathcal{T}$ given by the OIDS, the BN $\mathcal{B}$, and the new observation $(O_{new}, o_{new})$.

%obsolete observation deletion and substitution recommendation algorithm
\begin{algorithm}[!h]
\caption{Obsolete Observations Recommendation Algorithm (OORA)}
\begin{algorithmic}[1]
\label{algo1}
\renewcommand{\algorithmicrequire}{\textbf{Input:}}
\renewcommand{\algorithmicensure}{\textbf{Output:}}
\REQUIRE  $\mathcal{T}$, $\mathcal{B}$, $(O_{new}, o_{new})$
\ENSURE AND-OR recommendation tree \\
%\hspace{-0.58cm}\textbf{Parameters}: $\epsilon$: a real number, $0 \leq \epsilon \leq 1$\\

\STATE let $\textbf{S}_{o_{new}} = \{\textbf{S}_1,..., \textbf{S}_i,..., \textbf{S}_k\}$ the set regrouping all the subsets $\textbf{S}_i$ of the tree $\mathcal{T}$, such that $\textbf{S}_i= \{\textbf{S}_i^{AND}, \textbf{S}_i^{OR}\}$, $1 \leq k \leq N$

\FOR{each $\textbf{S}_i \in \textbf{S}_{o_{new}}$}
\STATE $\textbf{S}_i^{OR'} = \emptyset$ 
\STATE $\textbf{S}_i^{AND'} = \emptyset$ 

\FOR{each $(X, x) \in \textbf{S}_i^{OR}$}
\STATE $p_x = $ PosteriorProba($\mathcal{B}$, $x$, $o_{new}$)
\STATE $(x', p_{x'}) = $MostLikelyValue($X$, $\mathcal{B}$, $o_{new}$)
\STATE $\textbf{S}_i^{OR'} = \textbf{S}_i^{OR'} \cup (X, x, p_x, x', p_{x'})$
\ENDFOR
\STATE $\textbf{S}_i^{OR'} = Sort_{p_x}(\textbf{S}_i^{OR'})$

\FOR{each $(X, x) \in \textbf{S}_i^{AND}$}
\STATE $(x', p_{x'}) = $MostLikelyValue($X$, $\mathcal{B}$, $o_{new}$)
\STATE $\textbf{S}_i^{AND'} = \textbf{S}_i^{AND'} \cup (X, x, x', p_{x'})$
\ENDFOR

\STATE $\textbf{S}_i = \{\textbf{S}_i^{AND'}$, $\textbf{S}_i^{OR'}\}$
\STATE Update($\textbf{S}_{o_{new}}, \textbf{S}_i$)
\ENDFOR

\STATE UpdateTree($\mathcal{T}$, $\textbf{S}_{o_{new}}$)
\STATE \textbf{return} $\mathcal{T}$

\end{algorithmic}
\end{algorithm}

As a first step, for each $\textbf{S}_i \in \textbf{S}_{o_{new}}$ we calculate the posterior probability $p_x$ of each pair $(X,x) \in \textbf{S}_i^{OR}$, i.e., $P(x | o_{new})$, by calling the function \emph{PosteriorProba}.
Next, in order to help the target user make the right decisions, for each variable $X$ belonging to the set $\textbf{S}_i^{OR}$, we suggest the most likely value that can replace the older one taken  by $X$ by calling the function \emph{MostLikelyValue} (line 7). This function takes as input the variable $X$, the BN $\mathcal{B}$, the new observation $o_{new}$, predicts and returns the value $x'$ with the highest probability $p_{x'}$ that can replace the old value $x$ taken by $X$ by studying the posterior probability of $X$ giving $o_{new}$. This function is linear in the number of states of the variable $X$.

Let $\textbf{S}_i^{OR'}$ be the set of 5-tuples $(X, x, p_x, x', p_{x'})$ which represent each pair $(X, x)$ contained in the set $\textbf{S}_i^{OR}$ by associating it with the tuple $(p_x, x', p_{x'})$.
To prioritize the observations to be updated in the subset $\textbf{S}_i^{OR'}$ of each $\textbf{S}_i$, we sort them in ascending order of their posterior probabilities (line 10 of OORA).

We apply the same treatment on all $\textbf{S}_i^{OR}$ in $\textbf{S}_{o_{new}}$. For each $\textbf{S}_i^{OR'}$ resulted from line 10 of the OORA, the most likely observations to be updated are those with the lowest posterior probabilities, given $o_{new}$.
Line 11 traverses all the pairs $(X, x)$ in the subset $\textbf{S}_i^{AND} \in \textbf{S}_i$, and for each one, we call the function \emph{MostLikelyValue}.
The result from line 16 of our OORA is the set $\textbf{S}_{o_{new}} = \{\textbf{S}_1,..., \textbf{S}_k\}$ such that:

\begin{itemize}
  \item each $\textbf{S}_i^{AND}$ of each $\textbf{S}_i$ contains elements in the form of $(X, x, x',p_{x'})$, and
  \item each $\textbf{S}_i^{OR}$ of each $\textbf{S}_i$ contains elements in the form of $(X, x, p_x, x', p_{x'})$, and is sorted in ascending order of the calculated posterior probability $p_x$.
\end{itemize}

\begin{figure}[h]
  \centering
  \includegraphics[scale=0.7]{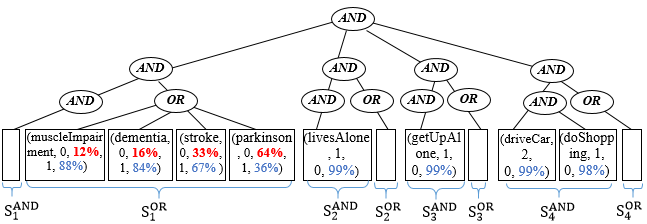}
  \caption{AND-OR recommendation tree associated with example \ref{example2}.} 
  \label{fig8}
\end{figure}

Once we have the new updated set $\textbf{S}_{o_{new}}$, our OORA update the AND-OR tree $\mathcal{T}$ based on $\textbf{S}_{o_{new}}$, and returns the AND-OR recommendation tree such that for each \emph{OR-Set}, the nodes with the lowest probability are packed from left to right. 

Our OORA is sound and complete and runs in $O(N_d \times N_x \times N_s)$, where $N_d$ is the size of the set $\textbf{S}_{o_{new}}$, $N_x$ is the number of pairs $(X, x)$ in $\textbf{S}_i$, and $N_s$ is the number of states of the variable $X$.

We apply the OORA on example \ref{example2}. Fig. \ref{fig8} shows the resulting AND-OR recommendation tree. In this example, we recommend checking observations on the variables \emph{livesAlone}, \emph{getUpAlone}, \emph{driveCar}, and \emph{doShopping} as they have become obsolete and require a mandatory update.  For the rest of observations, we recommend to start by checking the observation on the variable \emph{muscleImpairment} since it is the least probable given $o_{new}$.

\subsubsection{Predictions}
Now if we don't consider the contradictions, our RS aims to provide the caregivers with reliable information about their elderly patients in real-time (step 7 in Fig. \ref{fig2:system}) when some information is missing based on what it already knows about that older adult, denoted by $\textbf{OBS}$. Let $(X, x)$ denotes the requested information $x$ on the variable $X$ that carries the requested information. It consists in inferring the BN by observing $\textbf{OBS}$ and fixing the target variable to $X$, then returns the predicted information in the form of $(X, x, p_x)$ such that $p_x = P(X = x |\textbf{OBS})$ represents our belief about $X = x$ taking the observed data into account.
%\end{itemize}

\section{Empirical results}
\label{section5}

\subsection{Experimental setup}
%\subsubsection{Data}
As part of the Elderly-Fall Prevention project, we have access to the real-life database \emph{Elderly-Data} that contains information on the elderly gathered in both the University Hospital Center of Lille and Valenciennes. As previously reported, we used \emph{Elderly-Data}, among others, to build the causal BN for preventing falls among older adults. Once the Bayesian model is created and validated, we use it to generate a set of scenarios.
The generation of scenarios is done using an automatic generation process, which consists of (1) randomly selecting variables from the given BN, (2) assigning random observations to the selected variables, and (3) arbitrarily choosing a pair (variable, observation) that represents the newly acquired information. Thus each scenario $\textbf{S}_i$ is represented by the pair of newly acquired information (variable, new observation) accompanied by a sequence of pairs of some previously acquired observations that are consistent (variable, observed value). 
We generate a total of $700$ scenarios among which $580$ are chosen and validated by two experts, an orthopedist and a neurologist. For each scenario, $\textbf{S}_i$, a label $c_i$ given by the experts is associated such as: $c_i = 1$ if $\textbf{S}_i$ is declared contradictory by the experts, $c_i = 0$ otherwise. As a result, we obtain
%Our experiments are carried out using 
a balanced database $\mathcal{S}=\{(\textbf{S}_i, c_i)\}$ of $580$ scenarios labeled by experts with $290$ contradictory scenarios and $290$ non-contradictory scenarios that will be used to perform our experiments.

Then, for each scenario labeled as contradictory, the experts provided a list of subsets of all possible obsolete observations, such that the withdrawal of some observations of these subsets restores the consistency of the remaining observations with the newly acquired one. %The resulting subsets lists were then organized by experts into logical formulas. These formulas 
The resulting subsets will be compared later with the result provided by the OIDS.
Furthermore, for each subset of observations, the experts prioritize the obsolete observations and give us the most likely to be deleted. These observations will be compared later with those provided by our RS.

\subsection{Choice of the threshold $\epsilon$}

The detection of contradictory scenarios is conditioned by a threshold $\epsilon$. In most approximation-based works, the threshold is often hard to set. Various methods for parameter estimation can be proposed. In an ideal scenario, the value of $\epsilon$ can be set by a domain knowledge expert. In general, small values of $\epsilon$ are preferable. Alternatively, experimental studies and simulations on real-world databases can be applied to choose the optimum value of the threshold and thus minimize human intervention. In this work, we have applied the algorithm proposed by \cite{cbhd:data} on a part of the database $\mathcal{S}$ ($\approx35\%$ of the database) to calculate the $\epsilon$ value. It consists in varying the threshold over a sufficiently large range and tabulating the resulting true positive (TP) and false positive (FP) rates, then choosing the threshold that simultaneously guarantees better TP and FP rates. The threshold calculation algorithm used returns the value $10^{-2}$ relative to our BN for preventing falls in older adults.
Simultaneously, we have requested the assistance of our domain experts to validate the calculated threshold. They studied the BN, evaluated the resulting marginal distributions and how the distributions change after observing certain variables, tested the calculated threshold on some scenarios and validated its value, which is set at $10^{-2}$.

\subsection{Validation process}

The objective is to validate our OIUS in a real-life application, the elderly fall-prevention, and to showcase how the resulting  trees can be used to give reliable recommendations.
The validation consists of two parts: assess the quality of the AND-OR trees resulting from our OIDS and evaluate the quality of the recommendations provided by our RS.

\subsection*{\textbf{Step 1: Evaluation of the quality of the AND-OR trees}}
\paragraph{Precision}

We start by applying the OIDS on the rest of the database $\mathcal{S}$ ($\approx 65\%$ of the database) with $380$ scenarios divided into $190$ contradictory scenarios and $190$ non-contradictory scenarios labeled by experts (step 3 of Fig. \ref{fig2:system}). For each scenario, our OIDS estimates whether it is contradictory or not based on the BN and the given threshold.

\paragraph{Accuracy}

For scenarios classified as $\epsilon$-Contradictory by our system, we apply the obsolete observation identifying process (Step 5 in Fig. \ref{fig2:system}). As a result, we obtain %$185$
AND-OR trees, each relating to a contradictory scenario and encode all possible obsolete observations responsible for the contradiction.
To facilitate the assessment of the resulted AND-OR trees, we translate them, as well as the results provided by experts, into propositional formulas. Recall that the result of step 2 of the Obsolete information identification phase (Fig. \ref{fig4}) is the set $\textbf{S}_{o_{new}} = \{\textbf{S}_1, \textbf{S}_2, ..., \textbf{S}_k\}$. As explained in section \ref{section4}, when building the AND-OR tree, we group all the subsets $\textbf{S}_i$ under a root node labeled AND. This can be presented in the form of: $\textbf{S}_1 \land \textbf{S}_2 \land ... \land \textbf{S}_k$. Then, each subset $\textbf{S}_i$ is divided in two subsets $\textbf{S}_i^{AND}$ and $\textbf{S}_i^{OR}$, such that all observations of the set $\textbf{S}_i^{AND}$ (resp. $\textbf{S}_i^{OR}$) are linked with an \emph{AND} (resp. \emph{OR}). The formulas to be compared can therefore take the following form:
\[
\small \underbrace{(\underbrace{(a_1\land...\land a_i)}_{\textbf{S}_1^{AND}} \land \underbrace{(b_1\lor...\lor b_j)}_{\textbf{S}_1^{OR}})}_{\textbf{S}_1}\land...\land\underbrace{(\underbrace{(x_1\land...\land x_k)}_{\textbf{S}_p^{AND}}\land \underbrace{(y_1\lor...\lor y_l)}_{\textbf{S}_p^{OR}})}_{\textbf{S}_p}
\]

such that letters correspond to the possible obsolete observations and $i$, $j$, $k$, $l$, $p$ $\in \{1, ..., N\}$.
Subsequently, we compare each formula resulting from our OIDS with that given by the experts. 
The comparison is made in two levels. 
At the first level, we check the number of items (the subsets $\textbf{S}_i$) that appear in each formula.
%Out of the $185$ formulas relating to our OIDS, $185$ have the same number of $\textbf{S}_i$ as those given by the experts.
Then, for each proposition $\textbf{S}_i$, we compare the literals that form each of the clauses $\textbf{S}_i^{AND}$ and $\textbf{S}_i^{OR}$, with those provided by the experts.

\subsection*{\textbf{Step 2: Evaluation  of the quality of the recommendations}}

For the AND-OR trees resulting from the OIDS and that conform to those given by the experts, we apply the RS (Step 6 in Fig. \ref{fig2:system}). The results are AND-OR recommendation trees such that for each \emph{OR-Set} in the tree, the nodes that encodes the most likely obsolete observations are packed from left to right.
We compare each $\textbf{S}^{OR}_i$ in the tree resulting from our RS with the priority order assigned by the experts using a rank correlation that measures the relationship between different rankings of variables in the same set $\textbf{S}^{OR}_i$. Let $\textbf{S}^{OR}_i = \{X_1, X_2, ..., X_k\}$. We denote by $R$ and $S$ the assignment of the ordering labels '1', '2', '3', etc. to different variables $X_j$ in $\textbf{S}^{OR}_i$ assigned respectively by our RS and by the experts.
We denote by $r_j$ and $s_j$ the rankings of the variable $X_j$ assigned respectively by our RS and by the experts. Then we compare the two ranks $R$ and $S$ using the Spearman's rank correlation coefficient $\rho$ \cite{zar:significance} given by formula \ref{eq2}: 
%%% Kendall VS Spearman Rank Correlation !!

\begin{equation}
\label{eq2}
\rho = 1-{\frac {6\sum d_{j}^{2}}{k(k^{2}-1)}}
\end{equation}

where $d_j = r_j - s_j$ is the difference between the two rankings of each variable $X_j$ in $\textbf{S}^{OR}_i$ and $k$ is the number of variables in $\textbf{S}^{OR}_i$.

The Spearman's rank correlation coefficient can take values from $+1$ to $-1$. The $\rho$ is high when variables have a similar (or identical for a correlation of 1) ranking in both $R$ and $S$.
For each \emph{OR-Set}, we calculate the coefficient $\rho$ and we check if the priorities given by our RS match those suggested by the experts.

\subsection{Results and discussion}

Having chosen $10^{-2}$ as the appropriate threshold, we apply OIDA on the $380$ remaining scenarios.
The results obtained from Step 3 of Fig. \ref{fig2:system} are summarized in Table \ref{tab3}.
%out of the $190$ scenarios labeled by the experts as contradictory, $185$ scenarios are estimated with our algorithm as $\epsilon$-Contradictory and $5$ are considered $\epsilon$-non-contradictory. Out of the $190$ scenarios labeled by the experts as non-contradictory, $187$ are considered as $(1-\epsilon)$-non-contradictory by our algorithm and $3$ are considered $(1-\epsilon)$-Contradictory.

\begin{table}[h]
\begin{tabular}{lccc}
                                                      & \multicolumn{1}{l}{}                   & \multicolumn{2}{c}{\textbf{Predicted}}                                              \\ \cline{3-4} 
                                                      & \multicolumn{1}{c|}{N = 380}           & \multicolumn{1}{c|}{$\epsilon$-Contradictory} & \multicolumn{1}{c|}{$\epsilon$-non-contradictory} \\ \cline{2-4} 
\multicolumn{1}{l|}{\multirow{2}{*}{\textbf{Actual}}} & \multicolumn{1}{c|}{contradictory}     & \multicolumn{1}{c|}{185}               & \multicolumn{1}{c|}{5}                     \\ \cline{2-4} 
\multicolumn{1}{l|}{}                                 & \multicolumn{1}{c|}{non contradictory} & \multicolumn{1}{c|}{3}                 & \multicolumn{1}{c|}{187}                   \\ \cline{2-4} 
\end{tabular}
\caption{$10^{-2}$ Threshold contingency.}
\label{tab3}
\end{table}

Then, for the $185$ scenarios classified as $\epsilon$-Contradictory by our system, we apply the obsolete observation identifying process (Step 5 in Fig. \ref{fig2:system}).
Out of the $185$ propositional formulas relating to the AND-OR trees provided by our OIDS, $182$ are in line with those provided by the experts. Owing to space limitations, we cannot display all the results issuing from this step. However, some of the scenarios that were treated by our system are shown in Table \ref{fig8}.
Column 2 (resp. 3) of Table \ref{fig8} contains the newly (resp. previously) acquired observation(s). The observations underlined in column 3 are independent of the new observation and 
will not intervene in the rest of the treatment. 
%Column 4 refers to the step 1 of the obsolete information identification phase (Fig.\ref{fig4}), in which we seek to find among the previously acquired observations in column 3 those that depend on $O_{new}$. 
The conditional probabilities given in column 4 are all bellow the threshold $\epsilon$, which means that the scenarios presented are all contradictory. Column 5 (resp. 6) shows the results given by the OIDS (resp. experts). Note that for the first two scenarios in Table \ref{fig8}, the result provided by our OIDS is different from that given by the experts as it contains extra elements compared to that provided by the experts.

Next, we apply the RS on the $182$ AND-OR trees resulted from the OIDS that conform to those provided by the experts. We analyze the obtained $182$ recommendation trees from which a total of $273$ \emph{OR-Sets} are extracted. For each one, we calculate the Spearman's rank correlation coefficient $\rho$. For the $273$ \emph{OR-Sets}, the resulting coefficient $\rho$ ranges from $0.1$ to $1$ with an average of $0.73$ and is satisfactory for $232$ \emph{OR-Sets}.

We reviewed the results from the OIDS and focused on the three AND-OR trees that differ from those provided by the experts.
Tow of them are shown in Table \ref{fig8} and relate to the first two scenarios. In both scenarios 1 and 2, our OIDS presumes that observation on the variable \emph{psychotropicDrugs} may need to be checked and updated when $o_{new}$ is acquired, while the experts did not suggest it. For scenario 1, the emergence of this observation in the AND-OR tree provided by our OIDS comes from the fact that there is increasing evidence suggesting that psychotropic drugs used to treat psychiatric disorders could increase the risk of sudden cardiac death (SCD) \cite{sicouri:mechanisms}. Although identifying factors increasing the risk of SCD has been a challenge for both scientists and clinicians, there has been less information about the use of various drugs, such as psychotropic drugs, which may increase the vulnerability to fatal arrhythmias and this causal relationship remains a bit unpredictable.
For scenario 2, the results obtained from our OIDS can be explained by the fact that psychotropic drugs can induce movement disorders, in particular akinesia \cite{ward:antipsychotic}. This remains a bit tricky but important to know in clinical practice and requires additional research efforts on psychotropic-drugs side effects.
So we rechecked the result related to these two scenarios obtained from our OIDS with university hospital physicians. They confirmed the accuracy of the obtained results and appreciated the performance and precision of our system.

The error rate of our system comes from imperfections inherent in the given BN. Thus some conclusions will be incorrect, no matter how carefully drawn. Furthermore, a better representation cannot save us: all representations are imperfect, and any imperfection can be a source of error. Therefore we always make the assumption that all results are given with some degree of uncertainty. In this work, we have tested our approach on a BN that we built with some domain knowledge connoisseurs.
Note that our Bayesian model can be easily enhanced based on users feedback. Users feedback can be exploited to provide contextual details about errors or exceptions detected when using the OIUS; some variables can be added on-demand, some interactions can be added or deleted based on knowledge of local causal interactions with the existing parts and can be parameterized by expert judgment.

\begin{landscape}
\pagestyle{empty}
\begin{table}[h]
%\vspace{-2cm}
\hspace{-1.1cm}
\begin{tabular}{|p{1.5cm}|p{3cm}|p{4.5cm}|p{2cm}|p{3cm}|p{3cm}|p{1.5cm}|}
\hline
\textbf{Scenarios}  & $O_{new} = o_{new}$ & \thead{OBS'} & $P(O_{new}|\textbf{OBS'})$ & \thead{Results of OIDS} & \thead{Results of experts} & \textbf{Execution time} \\ 
\hline

\textbf{scenario 1} & cardiovascularDrugs = 1 & \{(heartDisease, 0), (drugsNb, 1), \underline{(akinesia, 0)}, (parkinson, 0), (diabetes, 0), (psychotropicDrugs, 0)\} & $3.3e^{-03}$  & 
\{heartDisease\} $\wedge$  \{drugsNb $\wedge$  psychotropicDrugs\}
& 
\{heartDisease\} $\wedge$  \{drugsNb\}& $7.28 ms$ \\ 
\hline

\textbf{scenario 2} & akinesia = 1 & \{(depression, 0), (psychotropicDrugs, 0), (parkinson, 0), (physiotherapy, 0), (driveCar, 2)\}  & $4.9e^{-05}$  & 
\{parkinson $\lor$ psychotropicDrugs\} $\wedge$ \{physiotherapy\}  $\wedge$ \{driveCar\} 
& 
\{parkinson\} $\wedge$ \{physiotherapy\}  $\wedge$ \{driveCar\} & $5.04 ms$ \\ 
\hline
\textbf{scenario 3} & fallsNb = 3 ($\ge$ 5 times) & \{(parkinson, 0), (strokeTIA, 0), (hypotension, 0), \underline{(diabetes, 1)}, (difficultyWalking, 0), (difficultyBalance, 0), \underline{(osteoporosis, 1)}, \underline{(muscleImpairment, 0)}, (getUpAlone, 0), (telealarm, 0)\} & $2.3e^{-05}$  
& 
\{strokeTIA $\lor$ hypotension $\lor$ difficultyWalking $\lor$ difficultyBalance\} $\wedge$ \{getUpAlone $\lor$ telealarm\} 
&
\{strokeTIA $\lor$ hypotension $\lor$ difficultyWalking $\lor$ difficultyBalance\} $\wedge$ \{getUpAlone $\lor$ telealarm\}  & $9.04 ms$ \\
\hline

\textbf{scenario 4} & autonomyLoss = 1   & \{(livesAlone, 1), (driveCar, 2), (parkinson, 0), \underline{(fearFalling, 0)}, (strokeTIA, 0), (difficultyWalking, 0), (muscleImpairment, 0)\}   & $2.1e^{-05}$     & 
\{muscleImpairment $\lor$ parkinson $\lor$ strokeTIA\} $\wedge$ \{driveCar\} $\wedge$ \{livesAlone\}
& 
\{muscleImpairment $\lor$ parkinson $\lor$ strokeTIA\} $\wedge$ \{driveCar\} $\wedge$ \{livesAlone\}  & $7.86 ms$ \\ 

\hline
\end{tabular}
\caption{An example of $10^{-2}$-Contradictory scenarios processing.}
\label{fig8}
\end{table}
\end{landscape}
\section{Conclusion}
\label{section6}
In this paper, we have outlined the design of a causal Bayesian model for preventing falls in older adults.
Identifying the elderly who are at risk of falling becomes a high - priority public health issue. Many healthcare services are now delivered in outpatient settings. During care, elderly assume significant responsibility for monitoring their own health status, managing their recovery, and communicating with clinicians from home. This distributed, mainly ambulatory, health care system enables older adults to participate in a proactive way, by regularly coordinating their care among multiple providers and by sharing health information with other third parties.
Since health care systems should be quick and easy to assess the fall risks and give some preventive recommendations, the elderly personal information collected and shared must be consistent and always up to date.
In this regard, we have proposed an obsolete personal information update system, which
aims to control and update in real-time the information acquired about each older adult, provide on-demand consistent information and supply tailored interventions to caregivers and fall-risk patients. The approach outlined for this purpose is based on a polynomial-time algorithm. The result is given as an AND-OR recommendation tree with some accuracy level. This tree encodes all possible obsolete observations. Moreover, it provides users with recommendations on which observations to delete and suggests the most likely values that can replace the obsolete ones. Therefore, it can be effectively used to maintain a consistent information base.
Analysis and simulations on a real database were conducted to evaluate the performance of the proposed approach.
Our approach efficiency is confirmed since results are very encouraging, reaching an accuracy of $93\%$.
In this paper, we present our model assuming that our BN is perfect and fixed during processing. Depending on the users' feedback and the error rate generated by our application, it can be revised after a while to meet the new requirements. We have assumed that the newly acquired information is certain. However, this may not be in real life, and this information may, therefore, make the recommendations inaccurate. So, in order to greatly expand the utility of our results, we will extend, in the future, our model to handle such different practical situations. One possible way is through uncertain evidence in the BN. In addition, a user interface will be proposed in order to perform a set of tests of our OIUS by some physicians, using an iterative and incremental development cycle.

\section*{Acknowledgements}
The present work is part of the ELSAT2020 project which is co-financed by the European Union with the European Regional Development Fund, the French state and the Hauts de France Region Council. It also relates to the "sustainable mobility and handicap" axis within the Polytechnic University of Hauts-de-France and is in collaboration with the university hospitals of Valenciennes and Lille. The experts who provided the estimates for the used BN and the University Hospital physicians who validated our scenarios are thanked for their participation.
\bibliographystyle{myelsarticle-num}
\bibliography{biblio}

\end{document}